\documentclass[preprint,showpacs,preprintnumbers,amsmath,amssymb,nofootinbib]{revtex4}

\usepackage{graphicx}
\usepackage{dcolumn}
\usepackage{bm}

\begin{document}
\title{\bf Time Delay Interferometry with Moving Spacecraft Arrays}

\author{Massimo Tinto}
\email{Massimo.Tinto@jpl.nasa.gov}
\altaffiliation [Also at: ]{Space Radiation Laboratory, California
  Institute of Technology, Pasadena, CA 91125}
\affiliation{Jet Propulsion Laboratory, California Institute of Technology, Pasadena, CA 91109}

\author{F. B. Estabrook}
\email{Frank.B.Estabrook@jpl.nasa.gov}
\affiliation{Jet Propulsion Laboratory, California Institute of Technology, Pasadena, CA 91109}

\author{J.W. Armstrong}
\email{John.W.Armstrong@jpl.nasa.gov}
\affiliation{Jet Propulsion Laboratory, California Institute of Technology,
 Pasadena, CA 91109}

\date{\today} 
\begin{abstract}
  
  Space-borne interferometric gravitational wave detectors, sensitive
  in the low-frequency (millihertz) band, will fly in the next decade.
  In these detectors the spacecraft-to-spacecraft light-travel-times
  will necessarily be unequal, time-varying, and (due to aberration)
  have different time delays on up- and down-links. Reduction of data
  from moving interferometric laser arrays in solar orbit will in fact
  encounter non-symmetric up- and downlink light time differences that
  are about $100$ times larger than has previously been recognized.
  The time-delay interferometry (TDI) technique uses knowledge of
  these delays to cancel the otherwise dominant laser phase noise and
  yields a variety of data combinations sensitive to gravitational
  waves.  Under the assumption that the (different) up- and downlink
  time delays are constant, we derive the TDI expressions for those
  combinations that rely only on four inter-spacecraft phase
  measurements. We then turn to the general problem that encompasses
  time-dependence of the light-travel times along the laser links. By
  introducing a set of non-commuting time-delay operators, we show
  that there exists a quite general procedure for deriving generalized TDI
  combinations that account for the effects of time-dependence of the
  arms. By applying our approach we are able to re-derive the
  ``flex-free'' expression for the unequal-arm Michelson
  combinations $X_1$, first presented in \cite{STEA}, and obtain the
  generalized expressions for the TDI combinations called Relay,
  Beacon, Monitor, and Symmetric Sagnac.

\end{abstract}

\maketitle

\section{Introduction}

Future space-borne gravitational wave (GW) observatories, such as LISA
\cite{PPA98}, will have sensitivity in the low-frequency band and will
use time-delay interferometry (TDI) to cancel laser phase noise.  All
the original papers on TDI considered a configuration of three
spacecraft interchanging coherent laser beams, and tacitly or
explicitly assumed the array to be at rest in an inertial system.  TDI
was treated in Euclidean 3-space with a universal time, in which the
velocity of light is $c$ and isotropic.  Recipes were given for
combining data (time series) separately recorded at the various
spacecraft, delayed by transit times calculated from the
inter-spacecraft separations $L_i$ (i = 1,2,3), in order to remove the
otherwise overwhelming phase noise of the laser sources
\cite{TA99,AET99,ETA00,TEA02,TAE01,TSSA,DNV}.  The aim is possible
detection of incident gravitational waves of galactic or cosmic
origin.

The LISA mission \cite{PPA98} will have three spacecraft orbiting the Sun
in a triangular array with the $L_i \simeq 5 \times 10^{6}$ km, and
GW detection capability in the band $10^{-4} - 1$ Hz. Several
TDI Michelson-like and Sagnac-like reduced laser-phase-noise-free data
streams will have different responses to secondary phase noise sources
and to two polarizations of incoming gravitational waves from
different directions.  A recent study of a linear array of three
spacecraft in a single solar orbit (SyZyGy) \cite{EATF} uses a TDI
combination sensitive to a single polarization of incident
gravitational waves, and two others sensitive solely to secondary
system noises.

In an important development, Shaddock \cite{S} noticed that rotational
motion of an array results in a difference of the light travel times
in the two directions around a Sagnac circuit.  Two time delays along
each arm must be used, say $L_i$ and $L_i^{'}$ for clockwise or
counterclockwise propagation as they enter in any of the TDI
combinations.  Shaddock emphasized the need for careful distinguishing
of primed and unprimed delays in the TDI combinations for
Michelson-like combinations, and, to eliminate laser noise from the
Sagnac-type combinations when the array is moving, he presented new
TDI variables related to those originally given by being ``double
differenced''.

Cornish and Hellings \cite{CH} also considered the effect of rotation
of the LISA triangle around its centroid on the TDI combinations, and
reported the new data combinations. Summers \cite{Summers} and Cornish
and Hellings \cite{CH} further pointed out that the LISA array is not
rigid, that $L_i$ and $L_i^{'}$ not only differ from one another but
can be time dependent (they "flex"), and that again the laser phase
noise (at least with present laser stability requirements) can enter
at a level above the secondary noises.  For LISA, and assuming $\dot
L_i \simeq 10 {\rm m/sec}$ \cite{Folkner}, they estimated the
magnitude of the remaining frequency fluctuations from the laser to be
about $30$ times larger than the level set by the secondary noise
sources in the center of the frequency band. This may not be as
serious a problem with SyZyGy \cite{EATF}.

Finally Shaddock et al. \cite{STEA} addressed the "flexing" complication
by showing that it becomes of higher order if the sequence of various
time delays in the new doubly differenced Sagnac combinations is
respected in the TDI recipe, and they introduced a new
doubly-differenced Michelson-type TDI combination to achieve the same
result.  They stressed that although all these combinations are
considerably more complicated than those originally given for a
non-moving array, and their GW response functions are similarly
complex, the final sensitivity $-$ calculated from GW signal strengths
and secondary phase noises $-$ is unaffected. 

All the analyses above, however, assumed the clocks onboard the three
LISA spacecraft to be synchronized to each other in a reference frame
attached to the LISA array. It is well known \cite{Ashby}, however,
that the spacetime geometry - here the Sagnac effect - prevents the
self-consistent synchronization of a network of clocks by transmission
of electromagnetic signals in a rotating reference frame. This implies
that the time adopted by the LISA onboard clocks and used for TDI has
to be referenced to an inertial reference frame and that the onboard
LISA receivers have to properly convert time information received from
Earth to the time in this inertial reference frame. Within this frame,
which we can assume to be Solar System Barycentric (SSB), the
differences between back-forth delay times that occur are in fact
thousands of kilometers, very much larger than has been previously
recognized by us or others. The problem is not rotation per se, but
rather aberration due to motion and changes of orientation in the SSB
frame.

In Section II, we further discuss the need for synchronizing the LISA
clocks with respect to a common inertial reference frame (SSB), and
the resulting GW response transfer functions.  We turn in Section
III to the derivation of the four-link TDI combinations valid for
constant time delays.  We first obtain the ``unequal-arm Michelson''
response, $X$, as an example of how time-delay operators can be used
for deriving TDI data combinations.  The operator formalism for TDI
was introduced by Dhurandhar {\it et al.}  \cite{DNV}. We use it in
conjunction with the usual subscripted delay notation to achieve a
systematic understanding of the ``Relay'' ($U, V, W$), ``Beacon'' ($P,
Q, R$), and ``Monitor'' ($E, F, G$) combinations. With laser
stabilization at a level somewhat improved from that used in the
original LISA study \cite{PPA98}, these combinations, now involving
different up- and down- link delays, will satisfy sensitivity
requirements.

In Section IV, however, we go on to use delay time/operator notation to
derive ``second generation'' TDI combinations, which account for
both the inequality and time dependence of the back/forth optical
paths. Following Shaddock et al.  \cite{STEA}, the resulting
doubly-differenced combinations, immune to first order shearing
(flexing, or constant rate of change of delay times) are denoted $X_i,
U_i, P_i, E_i \ , \ i = 1,2,3$. All these new combinations suppress
the nominal LISA laser phase noise to levels lower than those of the
secondary (proof-mass and optical-path) noise sources, and their
gravitational wave sensitivities are the same as previously computed
for the stationary case. For completeness, we calculate the remaining
shearing effect on the doubly-differenced versions of the
system-noise-monitoring combination $\zeta$, denoted $\zeta_1$,
$\zeta_2$, $\zeta_3$. Laser noise enters these combinations multiplied
by $\sin^2(\pi f L)$, where $f$ is a Fourier frequency in the LISA
band $10^{-4} - 1$ Hz.  We plot the laser noise in $\zeta_i$ for the
nominal LISA system and show it as a result also to be below the level of
secondary noises.

\section{Aberration, Time Delays and Synchronization of the LISA clocks}

The kinematics of the LISA and SyZyGy orbits brings in the effects of
motion at several orders of magnitude larger than any previous papers
on TDI have addressed. The instantaneous rotation axis of LISA, and
the SyZyGy array, both swing about the Sun at $30$ km/sec, and on any
leg the transit times of light signals in opposing directions, say
$L_i$ and $L_i^{'}$ ($c = 1$), can differ by as much as $1000$ km.
Aberration due to LISA's orbit about the Sun dominates its
instantaneous rotation. This observation reinforces the requirement
that the new TDI combinations of Section III and IV {\it{must}} be
used.  Indeed, $L_i$ and $L_i^{'}$ interchange periodically and so are
also time dependent; this effect is however of order $0.1$ m/sec and
is dominated by the effect of shearing (``flexing'') already
recognized.

This large motional effect has been overlooked because intuitively
up/down laser links between two spacecraft moving inertially on
parallel geodesics certainly appears symmetric in a co-moving frame.
The spacecraft are then seen ``at rest'' and the elapsed light times,
or delays, in either direction are the same. Consider however an
inertial frame in which two spacecraft are moving with speed $V$ along
a line, with constant separation.  The times of transit of a photon
from one to the other, forward or back, clearly must differ by
$2VL/c$.  This is just an extreme case of aberration.  There is no
paradox!  We have taken the speed of light to be $c$ and isotropic in
{\it{both}} frames, and special relativity has taught us that that is
fine so long as we properly re-synchronize the clocks that we use as
time coordinates when changing frame (using light beams!) The spatial
and temporal separations of two successive events along the null world
line of a ray of light {\it{ depend on choice of frame}}.  Rays
traveling in opposite senses between two moving spacecraft yield
different separations in all frames except the co-moving one.

An orbiting array is best described {\it{not}} by attempting a
sequence of co-moving tangent ``rest'' frames, but rather in the
barycentric non-rotating Euclidean frame moving with the Sun (of
course we ignore tiny general relativistic distortions).  The usual
time coordinate of positional solar system astronomy in principle uses
clocks such that c is isotropic. In LISA and SyZyGy data at the three
spacecraft will undoubtedly be taken and time-tagged in this Solar
System Barycentric frame, and all the up/down delay times used in the
new TDI combinations must be calculated from the coordinates of
emission and reception events in the SSB inertial frame. (This is
exactly parallel to the time synchronization problem, and its
resolution, that has been met by the designers of the GPS satellite
array \cite{Ashby} in geocentric orbit). 

Since the motion of the LISA array around the Sun introduces a
difference between (and a time dependence in) the co-rotating and
counter-rotating light travel times, the correct expressions for the
GW contributions to the various first-generation TDI combinations will
differ from the expressions valid for a stationary array \cite{AET99}.
The magnitude of the corrections introduced by the inequality of the
light-travel times is proportional to the product between the time
derivative of the GW amplitude and the differences between the actual
light travel times.  At one mHz, for instance, the correction to the
expression of the signal valid for a stationary array is five orders
of magnitude smaller.  Since the amplitude of this correction scales
linearly with the Fourier frequency, we can completely disregard this
effect (and also the weaker effect due to the time dependence of the
light travel times) over the entire LISA band.

It is clear, however, that over many months of continuous observation
of a quasi-periodic signal, the TDI responses have to account for the
motion of the array around the Sun (and relative to the GW source), 
which introduces secular modulations in the phase, frequency, and
amplitude of the GW responses \cite{C98}, \cite{CR03}.

\section{The Four-Link TDI combinations: Constant Time Delays}

The notation we will adopt is the same as used in the paper by
Shaddock {\it et al.}  \cite{STEA}, (i.e it is {\it{different}} from
the original TDI notation, e.g. Ref. \cite{ETA00}.)  We distinguish
time-of-flight delays by denoting with a prime those taken in the
counter-clockwise sense and unprimed delays in the clockwise sense
(see Fig. 1).

\begin{figure}
 \centering
\includegraphics[width=3.0 in, angle=0.0]{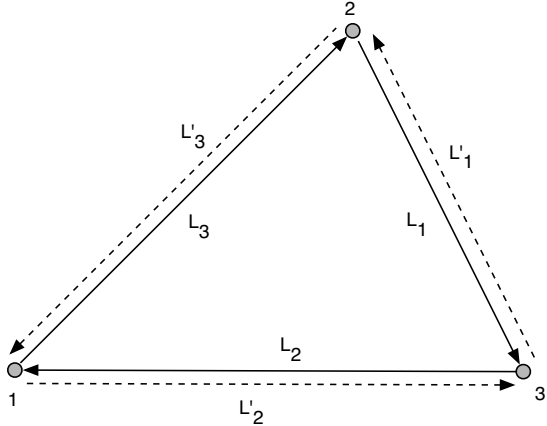}
\caption{Schematic diagram of LISA configurations involving
  six laser beams. Optical path delays taken in the counter-clockwise
  sense are denoted with a prime, while unprimed delays are in the
  clockwise sense. See text for details.}
\end{figure}

There are six beams exchanged between the LISA spacecraft, together
with the six phase measurements $s_{ij}$ ($i,j = 1, 2, 3$) recorded
when each transmitted beam is mixed with the laser light of the
receiving optical bench.  The phase fluctuations from the six lasers,
which need to be canceled, can be represented by six random processes
$p_{ij}$, where $p_{ij}$ is the phase of the laser in spacecraft $j$
on the optical bench facing spacecraft $i$. In what follows we assume
the center frequencies of the lasers are all equal, and denote it with
$\nu_0$.  Explicitly: $s_{23}$ is the one-way phase shift measured at
spacecraft $3$, coming from spacecraft $2$, along arm $1$.  The laser
phase noise in $s_{23}$ is $p_{32}(t - L_1) - p_{23}(t)$, where we
take $c = 1$, so that $L_1$ is the light time in the direction from
spacecraft $2$ to spacecraft $3$.  Similarly, $s_{32}$ is the phase
shift measured on arrival at spacecraft $2$ along arm $1'$ of a signal
transmitted from spacecraft $3$.  The laser phase noise in $s_{32}$ is
$p_{23}(t - L_1^{'}) - p_{32}(t)$, where $L_1^{'}$ is the light time
in the sense from 3 to 2 along arm $1^{'}$.  For the further delays
used in the TDI combinations we use the same conventions, being
careful to distinguish light travel along arms with primes or not,
depending on the sense of the measurement.  For example, our notation
for delaying the time series $s_{32}(t)$ by the clockwise light time
in arm 1 would be $s_{32,1}$ while delaying by the counterclockwise
light time in arm $1^{'}$ would be $s_{32,1'}$.  As before, we denote
six further data streams, $\tau_{ij}$ ($i,j = 1, 2, 3$), as the
intra-spacecraft metrology data used to monitor the motion of the two
optical benches and the relative phase fluctuations of the two lasers
on each of the three spacecraft.  The phase fluctuations of the lasers
and optical benches enter into the measurements $s_{ij}$ and
$\tau_{ij}$ in the following form \cite{TEA02} ({\it henceforth
  disregarding contributions from other noise sources and the
  gravitational wave signal})
\begin{eqnarray}
s_{31} & = &  \left[p_{13} - \nu_{0} \ {\hat n_2} \cdot {\vec \Delta_{13}} \right]_{,2} 
- \left[p_{31} + \nu_{0} \ {\hat n_2} \cdot {\vec \Delta_{31}}\right] \ ,
\label{eq:1}
\\
s_{21} & = & \left[p_{12} + \nu_{0} \ {\hat n_3} \cdot {\vec \Delta_{12}} \right]_{,3'}
- \left[p_{21} - \nu_{0} \ {\hat n_3} \cdot {\vec \Delta_{21}} \right] \ ,
\label{eq:2}
\\
\tau_{31} & = &  p_{21} - p_{31} - 2 \ \nu_{0} \ {\hat n_3} \cdot 
{\vec \Delta_{21}} + \mu_1 \ ,
\label{eq:3}
\\
\tau_{21} & = & p_{31} - p_{21} + \ 2 \ \nu_{0} \ {\hat n_2} \cdot {\vec
  \Delta_{31}} + \mu_1 \ .
\label{eq:4}
\end{eqnarray}
\noindent
In the above equations we have denoted with $\mu_i$ the phase
fluctuations introduced by the optical fibers used for exchanging the
laser beams between adjacent benches, and with the vector random
processes $\vec \Delta_{ij}$ the phase fluctuations introduced by the
mechanical vibrations of the optical benches.

In order to simplify the derivation of the new TDI combinations, we note
that by subtracting equation (\ref{eq:3}) from (\ref{eq:4}) we can
rewrite the resulting expression (and those obtained from it by
permutation of the spacecraft indices) in the following form
\begin{eqnarray}
{1 \over 2} \ [\tau_{21} - \tau_{31}] & = &
\left[p_{31} + \nu_{0} \ {\hat n_2} \cdot {\vec \Delta_{31}} \right]
- \left[p_{21} - \nu_{0} \ {\hat n_3} \cdot {\vec \Delta_{21}} \right]
\label{eq:5}
\\
{1 \over 2} \ [\tau_{32} - \tau_{12}] & = &
\left[p_{12} + \nu_{0} \ {\hat n_3} \cdot {\vec \Delta_{12}} \right]
- \left[p_{32} - \nu_{0} \ {\hat n_1} \cdot {\vec \Delta_{32}} \right]
\label{eq:6}
\\
{1 \over 2} \ [\tau_{13} - \tau_{23}] & = &
\left[p_{23} + \nu_{0} \ {\hat n_1} \cdot {\vec \Delta_{23}} \right]
- \left[p_{13} - \nu_{0} \ {\hat n_2} \cdot {\vec \Delta_{13}} \right]
\label{eq:7}
\end{eqnarray}
If we now define the following combinations of laser and optical bench noises
appearing in equations (\ref{eq:1}-\ref{eq:7}) \cite{DNV}
\begin{eqnarray}
\phi^*_{1} & \equiv & \left[p_{31} + \nu_{0} \ {\hat n_2} \cdot {\vec
    \Delta_{31}} \right] \ ,
\label{eq:8a}
\\
\phi_{1} & \equiv & \left[p_{21} - \nu_{0} \ {\hat n_3} \cdot {\vec
    \Delta_{21}} \right] \ ,
\label{eq:8b}
\end{eqnarray}
together with those obtained by permuting the spacecraft indices, it
is possible to reduce the derivation of the new TDI combinations to
the equivalent problem of removing the three random processes,
$\phi_1$, $\phi_2$, and $\phi_3$, from the following six linear
combinations of the one-way measurements $s_{ij}$ and $\tau_{ij}$:
\begin{eqnarray}
\eta_{21}  & \equiv &  s_{21} - {1 \over 2} \ [\tau_{32} - \tau_{12}]_{,3'} 
= \phi_{2,3'} - \phi_1
\ , \ 
\eta_{31} \equiv s_{31} + {1 \over 2} \ [\tau_{21} - \tau_{31}] 
= \phi_{3,2} - \phi_1 \ ,
\label{eq:9a}
\\
\eta_{12} & \equiv & s_{12} + {1 \over 2} \ [\tau_{32} - \tau_{12}]
= \phi_{1,3} - \phi_2 
\ , \ \ \ \ 
\eta_{32} \equiv s_{32} - {1 \over 2} \ [\tau_{13} - \tau_{23}]_{,1'}
= \phi_{3,1'} - \phi_2 \ ,
\label{eq:9b}
\\
\eta_{13} & \equiv & s_{13} -  {1 \over 2} \ [\tau_{21} - \tau_{31}]_{,2'}
= \phi_{1,2'} - \phi_3
\ , \ 
\eta_{23} \equiv s_{23} + {1 \over 2} \ [\tau_{13} - \tau_{23}]
= \phi_{2,1} - \phi_3 \ .
\label{eq:9c}
\end{eqnarray}

\subsection{The Unequal-Arm Michelson}

Here we derive the unequal-arm Michelson combination, $X$, valid for
the rigid-rotation case.  We use $X$ as an example for deriving
TDI data combinations by using an alternative and powerful method
based on the use of properly defined time-delay operators.

The $X$ combination relies on the four measurements $\eta_{12}$,
$\eta_{21}$, $\eta_{13}$, and $\eta_{31}$. Note that the two
combinations $\eta_{21} + \eta_{12,3'}$, $\eta_{31} + \eta_{13,2}$,
which represent the two synthesized two-way data measured onboard
spacecraft $1$, can be written in the following form
\begin{eqnarray}
\eta_{21} + \eta_{12,3'} & = & \left(D_{3'}D_{3} - I\right) \ \phi_1 \ ,
\label{eq:10a}
\\
\eta_{31} + \eta_{13,2}  & = & \left(D_{2}D_{2'} - I\right) \ \phi_1 \ ,
\label{eq:10b}
\end{eqnarray}
where we have denoted with $D_j$ the time-delay operator that shifts
by $L_j$ the function it is applied to, and with $I$ the identity
operator. Note that in the stationary case any pairs of these
operators commute, i.e.  $D_i D_{j'} - D_{j'} D_i = 0$ (while they do
not when the delays are functions of time \cite{CH}, \cite{STEA}).

\noindent
From equations (\ref{eq:10a}, \ref{eq:10b}) it is easy to derive the
following expression for $X$, by requiring the elimination of $\phi_1$
\begin{eqnarray}
X & = & \left[D_{2}D_{2'} - I\right] (\eta_{21} + \eta_{12,3'}) - 
\left[\left(D_{3'}D_{3} - I\right)\right] (\eta_{31} + \eta_{13,2})
\nonumber \\
& = & [(\eta_{31} + \eta_{13,2}) + (\eta_{21} + \eta_{12,3'})_{,2'2}]
- [(\eta_{21} + \eta_{12,3'}) + (\eta_{31} + \eta_{13,2})_{,33'}]
\label{eq:11}
\end{eqnarray}
After replacing equations (\ref{eq:9a}, \ref{eq:9b}, \ref{eq:9c}) into
equation (\ref{eq:11}), we obtain the final expression for $X$ valid
in the case of rigid rotation of the LISA array \cite{S}
\begin{eqnarray}
X & = & [(s_{31} + s_{13,2}) + (s_{21} + s_{12,3'})_{,2'2}]
- [(s_{21} + s_{12,3'}) + (s_{31} + s_{13,2})_{,33'}]
\nonumber \\
& & + {1 \over 2} [(\tau_{21} - \tau_{31})_{,2'233'} - (\tau_{21} -
\tau_{31})_{,33'} - (\tau_{21} - \tau_{31})_{,2'2} + (\tau_{21} - \tau_{31})]
\label{eq:12a}
\end{eqnarray}
\noindent

\noindent
As pointed out in \cite{Summers} and \cite{STEA}, equation
(\ref{eq:11}) shows that $X$ is the difference of two sums of phase
measurements, each corresponding to a specific light path from a laser
onboard spacecraft $1$ having phase noise $\phi_1$. The first
square-bracket term in equation (\ref{eq:11}) represents a synthesized
light-beam transmitted from spacecraft $1$ and made to bounce once at
spacecraft $3$ and $2$ respectively. The second square-bracket term
instead correspond to another beam also originating from the same
laser, experiencing the same overall delay as the first beam, but
bouncing off spacecraft $2$ first and then spacecraft $3$. When they
are recombined they will cancel the laser phase fluctuations exactly,
having both experienced the same total delays (assuming stationary
spacecraft).

\subsection{The Relay}

The TDI ``Relay'' configurations were called ($U, V, W$) (equation
(A4) of \cite{ETA00}).  In what follows, let us consider, as a
specific example, the $U$ combination, which has to rely only on the
four measurements $\eta_{31}$, $\eta_{12}$, $\eta_{32}$ and
$\eta_{23}$. The idea we will follow for identifying the expression
for $U$ is to select combinations of some of these four measurements
that contain only one phase noise. By then applying iteratively the
time-delay procedure we introduced for the $X$ combination, we will be
able to remove all the phase noises $\phi_i  , i=1, 2, 3$.  Note
that the obvious combinations that contain only one of the three phase
noises $\phi_i$ are the synthesized two-way Doppler data measured
onboard spacecraft $2$ and $3$. They in fact contain only the phase
noises $\phi_2$ and $\phi_3$ respectively.  Since the remaining two
measurements $\eta_{12}$ and $\eta_{31}$ can be combined in such a way
as to eliminate the phase noise $\phi_1$, we can start with the following
set of three data combinations
\begin{eqnarray}
\eta_{12} + \eta_{31,3} & = & D_{3} D_{2} \phi_3 - \phi_2 \ ,
\label{eq:19a}
\\
\eta_{32,1} + \eta_{23} & = & \left[D_{1}D_{1'} - I\right] \phi_3 \ ,
\label{eq:19b}
\\
\eta_{23,1'} + \eta_{32} & = & \left[D_{1'}D_{1} - I\right] \phi_2 \ .
\label{eq:19c}
\end{eqnarray}
It is then easy to see that the expression for $U$ is given by
the following linear combination of the properly delayed equations
(\ref{eq:19a}, \ref{eq:19b}, \ref{eq:19c})
\begin{eqnarray}
U & = & \left[D_{1'}D_{1} - I\right] \left(\eta_{12} + \eta_{31,3}\right)
+ \left(\eta_{23,1'} + \eta_{32}\right) - 
D_3 D_2 \left(\eta_{32,1} + \eta_{23}\right) \ ,
\nonumber
\\
& = & \left(\eta_{12,11'} + \eta_{31,311'}\right) - \left(\eta_{12} +
\eta_{31,3}\right) + \left(\eta_{23,1'} + \eta_{32}\right) - \left(\eta_{32,123} + \eta_{23,23}\right) \ ,
\label{eq:20}
\end{eqnarray}
which, in terms of the one-way measurements $s_{ij}$ and $\tau_{ij}$, becomes
\begin{eqnarray}
U & = & s_{31,311'} - s_{31,3} + s_{12,11'} - s_{12} + s_{23,1'} 
+ s_{32} - s_{32,123} - s_{23,23} 
\nonumber \\
& & + {1 \over 2} [(\tau_{21} - \tau_{31})_{,311'} - (\tau_{21} -
\tau_{31})_{,3} - (\tau_{32}- \tau_{12}) + (\tau_{32}- \tau_{12})_{,11'}
\nonumber \\
& & + (\tau_{13}- \tau_{23})_{,1'123} - (\tau_{13}- \tau_{23})_{,23}]
\label{eq:21}
\end{eqnarray}
\noindent
with $V$, $W$ obtained by cycling the spacecraft indices.

\subsection{The Beacon}

In the ``Beacon'' combination, one spacecraft transmits (only) to the
other two while those other two exchange one-way beams as usual.
These were called the ($P,Q,R$) combinations, depending on which
spacecraft was the transmit-only element of the array \cite{ETA00}. In
order to derive the expression for $P$, which involves only the four
data streams $\eta_{12}$, $\eta_{13}$, $\eta_{32}$, and $\eta_{23}$, we
will proceed according to the above considerations, and use in
this case the following data combinations
\begin{eqnarray}
\eta_{12,2'} - \eta_{13,3} & = & D_{3} \phi_3 - D_{2'} \phi_2 \ ,
\label{eq:13a}
\\
\eta_{32,1} + \eta_{23} & = & \left[D_{1}D_{1'} - I\right] \phi_3 \ ,
\label{eq:13b}
\\
\eta_{23,1'} + \eta_{32} & = & \left[D_{1'}D_{1} - I\right] \phi_2 \ .
\label{eq:13c}
\end{eqnarray}
By taking advantage of the commutativity of the delay operators in
this constant time delay case, it is easy to see that the expression
for $P$ is given by the following linear combination of the properly
delayed equations (\ref{eq:13a}, \ref{eq:13b}, \ref{eq:13c})
\begin{eqnarray}
P & = & D_{3} \left(\eta_{32,1} + \eta_{23}\right) - 
D_{2'} \left(\eta_{23,1'} + \eta_{32}\right) - 
\left[D_{1'}D_{1} - I\right] \left(\eta_{12,2'} - \eta_{13,3}\right) 
\ ,
\nonumber
\\
& = & \left(\eta_{32,13} + \eta_{23,3}\right) - 
\left(\eta_{23,1'2'} + \eta_{32,2'}\right) -  \left(\eta_{12,2'11'} - \eta_{13,311'}\right) 
+ \left(\eta_{12,2'} - \eta_{13,3}\right) \ .
\label{eq:14}
\end{eqnarray}
Equation (\ref{eq:14}) can be rewritten in terms of the one-way
measurements $s_{ij}$, $\tau_{ij}$
\begin{eqnarray}
P & = & s_{12,2'} - s_{13,3} - s_{32,2'} + s_{23,3} +
s_{32,13} - s_{23,1'2'} + s_{13,311'} - s_{12,2'11'}
\nonumber \\
& & + {1 \over 2} [(\tau_{21} - \tau_{31})_{,2'3} - (\tau_{21} -
\tau_{31})_{,11'2'3} + (\tau_{32}- \tau_{12})_{,2'} - (\tau_{32}- \tau_{12})_{,11'2'}
\nonumber \\
& & + (\tau_{13}- \tau_{23})_{,3} - (\tau_{13}- \tau_{23})_{,11'3}] \ ,
\label{eq:15}
\end{eqnarray}
\noindent
with $Q$, $R$ obtained by cycling the spacecraft indices in Eq.
(\ref{eq:15}).

\subsection{The Monitor}

Similarly, there are three combinations where one spacecraft is
listen-only \cite{ETA00}.  In order to derive these ``Monitor''
combinations ($E, F, G$) (equation (A1) of \cite{ETA00}), let us
consider the following combinations of the four data streams that
enter into $E$
\begin{eqnarray}
\eta_{31} - \eta_{21} & = & D_{2} \phi_3 - D_{3'} \phi_2 \ ,
\label{eq:16a}
\\
\eta_{32,1} + \eta_{23} & = & \left[D_{1}D_{1'} - I\right] \phi_3 \ ,
\label{eq:16b}
\\
\eta_{23,1'} + \eta_{32} & = & \left[D_{1'}D_{1} - I\right] \phi_2 \ .
\label{eq:16c}
\end{eqnarray}
Similarly to the derivations made for the two previous combinations,
it is easy to see that the expression for
$E$ is given by the following linear combination of the properly
delayed equations (\ref{eq:16a}, \ref{eq:16b}, \ref{eq:16c})
\begin{eqnarray}
E & = & D_{2} \left(\eta_{32,1} + \eta_{23}\right) - 
D_{3'} \left(\eta_{23,1'} + \eta_{32}\right) - 
\left[D_{1'}D_{1} - I\right] \left(\eta_{31} - \eta_{21}\right) \ ,
\nonumber
\\
& = & \left(\eta_{32,12} + \eta_{23,2}\right) - 
\left(\eta_{23,1'3'} + \eta_{32,3'}\right) -  
\left(\eta_{31,11'} - \eta_{21,1'1}\right) + 
\left(\eta_{31} - \eta_{21}\right) \ ,
\label{eq:17}
\end{eqnarray}
which, in terms of the one-way measurements $s_{ij}$ and $\tau_{ij}$ becomes
\begin{eqnarray}
E & = & s_{32,12} + s_{23,2} - s_{23,1'3'} - s_{32,3'} 
+ s_{21,1'1} - s_{31,11'} - s_{21} + s_{31}
\nonumber \\
& & + {1 \over 2} [(\tau_{21} - \tau_{31}) - (\tau_{21} -
\tau_{31})_{,11'} + (\tau_{32}- \tau_{12})_{,3'} - (\tau_{32}- \tau_{12})_{,11'3'}
\nonumber \\
& & + (\tau_{13}- \tau_{23})_{,2} - (\tau_{13}- \tau_{23})_{,11'2}]
\label{eq:18}
\end{eqnarray}
\noindent
with $F$, $G$ obtained by cycling the indices.

\subsection{The $\zeta$ Combinations}

In all the above, we have used the same symbol (e.g., $X$ for the
unequal-arm Michelson combination) for both the rotating (i.e.
constant delay times) and stationary cases.  This emphasizes that, for
these TDI combinations, the forms of the equations do not change going
from systems at rest to the moving or rotating case.  One need only
distinguish between the time-of-flight variations in the clockwise and
counter-clockwise senses (primed and unprimed delays).

In the case of an array at rest there is one symmetric data
combination that cancels exactly all laser noise and optical bench
motions and has the property that each of the $\eta_{ij}$ enters exactly
once and is lagged by exactly one of the one-way light times.  We
called this $\zeta$ (\cite{ETA00}, equation (3.5)) and showed how to
take advantage of its relative immunity to GWs in order to assess
on-orbit instrumental noise performance and distinguish instrumental
noise from a confusion-limited background \cite{TAE01}. Although now
the rotation of the array breaks the symmetry and therefore the
uniqueness of a ``$\zeta$-like'' combination, it has been shown
(\cite{S}, \cite{CH}) that there still exist three generalized TDI
laser-noise-free data combinations that have properties very similar
to $\zeta$, and which can be used for the same scientific purposes.
Here we derive these combinations, which we call ($\zeta_1, \zeta_2,
\zeta_3$), by applying our time-delay operator approach. As we will
see in the following section, our derivation will automatically
identify the ``correct'' order of the delays that has to be applied
to the one-way data. In other words, the expressions lead to an order
of time delays such that even with shearing the remaining laser noise
is below the level identified by the secondary noise sources.
$\zeta_1$ will not have to be further generalized. 

\noindent
Let us consider the following combination of the $\eta_{ij}$
measurements
\begin{eqnarray}
\eta_{13,3'} - \eta_{23,3'} + \eta_{21,1} & = & \left[D_{3'} D_{2'} -
  D_{1}\right] \phi_1 \ ,
\label{eq:22a}
\\
\eta_{31,1'} - \eta_{32,2} + \eta_{12,2} & = & \left[D_{3} D_{2} -
  D_{1'}\right] \phi_1 \ ,
\label{eq:22b}
\end{eqnarray}
where we have used the commutativity property of the delay operators
in order to cancel the $\phi_2$ and $\phi_3$ terms. Since both sides
of the two equations above contain only the $\phi_1$ noise, 
$\zeta_1$ is found by the following expression
\begin{equation}
\zeta_1 = \left[D_{3'}D_{2'} - D_{1}\right] \left(\eta_{31,1'} - \eta_{32,2} + \eta_{12,2}\right)
- \left[D_{2}D_{3} - D_{1'}\right]\left(\eta_{13,3'} - \eta_{23,3'} +
  \eta_{21,1}\right) \ .
\label{eq:23}
\end{equation}
In terms of the one-way measurements $s_{ij}$ and $\tau_{ij}$,
equation (\ref{eq:23}) becomes
\begin{eqnarray}
\zeta_1 & = & [s_{31,1'} - s_{32,2} + s_{12,2}]_{,2'3'} -
[s_{13,3'} - s_{23,3'} + s_{21,1}]_{,32} 
\nonumber \\
& & - [s_{31,1'} - s_{32,2} + s_{12,2}]_{,1} 
+ [s_{13,3'} - s_{23,3'} + s_{21,1}]_{,1'}
\nonumber \\
& & + {1 \over 2} [(\tau_{32} - \tau_{12})_{,22'3'} 
- (\tau_{32} - \tau_{12})_{,21} + (\tau_{32} - \tau_{12})_{,13'32} 
- (\tau_{32} - \tau_{12})_{,13'1'} 
\nonumber \\
& & + (\tau_{13}- \tau_{23})_{,22'3'1'} - (\tau_{13}- \tau_{23})_{,211'}
+ (\tau_{13}- \tau_{23})_{,233'} - (\tau_{13}- \tau_{23})_{,3'1'}
\nonumber \\
& & + (\tau_{21}- \tau_{31})_{,22'33'} - (\tau_{21}-
\tau_{31})_{,11'}]
\label{eq:24}
\end{eqnarray}
\noindent
together with its cyclic permutations.  (This expression for $\zeta_1$
was given (but not derived) in \cite{S} and independently by
\cite{CH}.)  If the light-times in the arms are equal in the clockwise
and counterclockwise senses (e.g. no rotation) there is no distinction
between primed and unprimed delay times.  In this case, $\zeta_1$ is
related to our original symmetric Sagnac $\zeta$ by $\zeta_1 =
\zeta_{,23} - \zeta_{,1}$.  Thus for the practical LISA case (arm
length difference $< 2 \%$), the SNR of $\zeta_1$ will be the same as
the SNR of $\zeta$.

\section{The Second-Generation TDI combinations}

Generalizations of the original unequal-arm Michelson, ($X, Y, Z$),
and Sagnac, ($\alpha, \beta, \gamma$) TDI combinations to an array
with systematic spacecraft velocities, showing that they effectively
cancel all laser phase noises, have been derived in \cite{STEA}. Here
we complete that set of TDI combinations by deriving generalized
expressions for the ``Relay'', ``Beacon'', and ``Monitor''
combinations that are unaffected by the rotation and time-dependence
of the light-path delays. These TDI combinations rely only on four of
the six possible one-way measurements LISA will make, and for this
reason they add robustness and trade-off options to the LISA design.
Like the unequal-arm Michelson combination $X_1$ \cite{STEA}, these
new combinations involve the four one-way inter-spacecraft
measurements at $16$ different times.

The {\it{order}} of the time-delay operators now becomes important for
laser phase terms.  The operators can no longer be permuted freely to
show cancellation of laser noises in the TDI combinations (they no
longer commute!). In order to derive the new, ``flex-free'' Relay,
Beacon, and Monitor combinations we will start by taking specific
combinations of the one-way data entering in each of the expressions
derived in the previous section for the rigid-rotation case. These
combinations are chosen in such a way to retain only one of the three
noises $\phi_i , i=1, 2, 3$ if possible. In this way we can then
implement an iterative procedure based on the use of these basic
combinations and of time-delay operators, to cancel the laser noises
after dropping terms that are quadratic in $\dot{L}/c$ or linear in
the accelerations. This iterative time-delay method, to first order in
the velocity, is illustrated abstractly as follows.  Given a function
of time $\Psi = \Psi(t)$, time delay by $L_i$ is now denoted either
with the standard comma notation or by applying the delay operator
$D_{i}$ introduced in the previous section
\begin{equation}
D_{i} \Psi = \Psi_{,i} \equiv \Psi(t - L_i(t)) 
\label{eq:25}
\end{equation}
\noindent
We then impose a second time delay $L_j(t)$:
\begin{eqnarray}
D_{j} D_{i} \Psi = \Psi_{;ij} & \equiv &  \Psi(t - L_j(t) - L_i(t - L_j(t)))
\nonumber \\
&  \simeq  & \Psi(t - L_j(t) - L_i(t) + \dot  L_i(t)  L_j) 
\nonumber \\
&  \simeq & \Psi_{,ij} + \dot \Psi_{,ij} \dot L_i L_j
\label{eq:26}
\end{eqnarray}

\noindent
A third time delay $L_k(t)$ gives:
\begin{eqnarray}
D_{k} D_{j} D_{i} \Psi = \Psi_{;ijk} & = & 
\Psi(t - L_k(t) - L_j(t - L_k(t)) - L_i(t - L_k(t) - L_j(t - L_k(t))))
\nonumber \\
& & \simeq \Psi_{,ijk} + \dot \Psi_{,ijk} [\dot L_i (L_j + L_k) + \dot L_j L_k]
\label{eq:27}
\end{eqnarray}
\noindent
and so on, recursively; each delay generates a first-order correction
proportional to its rate of change times the sum of all delays coming
after it in the subscripts. Commas have now been replaced with
semicolons \cite{STEA}, to remind us that we consider moving arrays.
When the sum of these corrections to the terms of a data combination
vanishes, the combination is called flex-free.

Also, note that each delay operator, $D_{i}$, has a unique inverse,
$D^{-1}_{i}$, whose expression can be derived by requiring that
$D^{-1}_{i} D_{i} = I$, and neglecting quadratic and higher order
velocity terms. Its action on a time series $\Psi (t)$ is
\begin{equation}
D^{-1}_{i} \Psi (t) \equiv \Psi (t + L_i (t + L_i)) \ .
\label{eq:28}
\end{equation}
\noindent
Note that this is not like an advance operator one might expect,
since it advances not by $L_i (t)$ but rather $L_i (t + L_i)$.

\subsection{The Unequal-Arm Michelson}

Here we re-derive the generalized unequal-arm Michelson combination
\cite{STEA}, $X_1$, by implementing our method based on the use of
time-delay operators. We use again $X_1$ as an example for showing the
effectiveness of this alternative and powerful method for deriving TDI
data combinations accounting for rotation and time-dependence of the
LISA arms.

Let us consider the following two combinations of the one-way measurements
entering into the $X$ observable given in the previous section,
evaluating them for the noises $\phi_i$ only
(equation \ref{eq:11})
\begin{eqnarray}
\left[(\eta_{31} + \eta_{13;2}) + (\eta_{21} + \eta_{12;3'})_{;2'2}\right]  & = &  
\left[D_{2}D_{2'}D_{3'}D_{3} - I \right] \phi_1 \ ,
\label{eq:29a}
\\
\left[(\eta_{21} + \eta_{12;3'}) + (\eta_{31} + \eta_{13;2})_{;33'}\right] & = &
\left[D_{3'}D_{3}D_{2}D_{2'} - I \right] \phi_1 \ .
\label{eq:29b}
\end{eqnarray}
If the time delays were constants, so the operators on the right would
permute freely, simply differencing of equations (\ref{eq:29a},
\ref{eq:29b}) eliminates $\phi_1$ and indeed is just $X$. If they do
not permute, from equations (\ref{eq:29a}, \ref{eq:29b}) we can use
the delay technique again to write the following expression for $X_1$
\begin{eqnarray}
X_1 & = & \left[D_{2}D_{2'}D_{3'}D_{3} - I \right] \ 
\left[(\eta_{21} + \eta_{12;3'}) + (\eta_{31} +
  \eta_{13;2})_{;33'}\right]
\nonumber
\\
& & -
\left[D_{3'}D_{3}D_{2}D_{2'} - I \right] \ 
\left[(\eta_{31} + \eta_{13;2}) + (\eta_{21} + \eta_{12;3'})_{;2'2}\right]
\nonumber \\
& = & \left[(\eta_{31} + \eta_{13;2}) + (\eta_{21} +
  \eta_{12;3'})_{;2'2} + (\eta_{21} + \eta_{12;3'})_{;33'2'2} + (\eta_{31} +
  \eta_{13;2})_{;33'33'2'2} \right]
\nonumber \\
& & -  \left[(\eta_{21} + \eta_{12;3'}) + (\eta_{31} +
  \eta_{13;2})_{;33'} + (\eta_{31} + \eta_{13;2})_{;2'233'} + 
(\eta_{21} + \eta_{12;3'})_{;2'22'233'} \right] \ .
\label{eq:30}
\end{eqnarray}
After substituting equations (\ref{eq:9a}, \ref{eq:9b}, \ref{eq:9c})
into equation (\ref{eq:30}), we obtain the final expression
for $X_1$ \cite{STEA}
\begin{eqnarray}
X_1 & = & [(s_{31} + s_{13;2}) + (s_{21} + s_{12;3'})_{;2'2}
+ (s_{21} + s_{12;3'})_{;33'2'2}
+ (s_{31} + s_{13;2})_{;33'33'2'2}]
\nonumber
\\
& & -
[(s_{21} + s_{12;3'})
+ (s_{31} + s_{13;2})_{;33'}
+ (s_{31} + s_{13;2})_{2'233'} +
(s_{21} + s_{12;3'})_{;2'22'233'}]
\nonumber
\\
& & +  {1 \over 2} \ [(\tau_{21} - \tau_{31}) - (\tau_{21} -
\tau_{31})_{;33'} - (\tau_{21} - \tau_{31})_{;2'2}
+ (\tau_{21} - \tau_{31})_{;33'33'2'2}
\nonumber
\\
& & + (\tau_{21} - \tau_{31})_{;2'22'233'}
- (\tau_{21} - \tau_{31})_{;2'233'33'2'2}] \ ,
\label{eq:31}
\end{eqnarray}

\noindent
As usual, $X_2$ and $X_3$ are obtained by cyclic permutation of the
spacecraft indices. This expression is readily shown to be laser-noise-free
to first order of spacecraft separation velocities $\dot L_i$: it is
``flex-free''.

\subsection{The Relay}

In order to derive the expressions for the generalized Relay
combinations ($U_1, U_2, U_3$) valid for the realistic kinematical
configuration of the LISA spacecraft, let us consider the following
combinations of the data that enter into the expression for $U$ given
in the previous section
\begin{eqnarray}
\left[ \eta_{12} + \eta_{31;3} + \eta_{23;23} + \eta_{32;123} - \eta_{32} \right]
& = & \left[D_{3}D_{2}D_{1} - I \right] D_{1'} \phi_3 \ ,
\label{eq:35a}
\\
\left[ \eta_{12;11'} +  \eta_{23;1'} +  \eta_{31;311'}\right]
 & = & D_{1'} \left[D_{1}D_{3}D_{2} - I \right] \phi_3 \ .
\label{eq:35b}
\end{eqnarray}
In each case we evaluate them for the noises $\phi_i$ only, as these
are what our combinations must remove.  The expression for $U_1$ is
then given by the following linear combination of the properly delayed
equations (\ref{eq:35a}, \ref{eq:35b})
\begin{eqnarray}
U_1 & = & D_{1'} \left[D_{1}D_{3}D_{2} - I \right]
\left[ \eta_{12} + \eta_{31;3} + \eta_{23;23} + \eta_{32;123} - \eta_{32} \right]
\nonumber
\\
& & - \left[D_{3}D_{2}D_{1} - I \right] D_{1'}
\left[ \eta_{12;11'} +  \eta_{23;1'} +  \eta_{31;311'}\right]
\label{eq:36}
\\
& = & \left[ \eta_{12;2311'} + \eta_{31;32311'} + \eta_{23;232311'} +
  \eta_{32;1232311'} - \eta_{32;2311'} \right]
\nonumber
\\
& & - \left[ \eta_{12;1'} + \eta_{31;31'} + \eta_{23;231'} +
  \eta_{32;1231'} - \eta_{32;1'} \right]
\nonumber
\\
& & + \left[ \eta_{12;11'1'} +  \eta_{23;1'1'} +
  \eta_{31;311'1'}\right]
- 
\left[ \eta_{12;11'1'123} +  \eta_{23;1'1'123} +
  \eta_{31;311'1'123}\right]
\label{eq:36a}
\end{eqnarray}
which, in terms of the one-way measurements $s_{ij}$ and $\tau_{ij}$, becomes
\begin{eqnarray}
U_1 & = & \left[ s_{12;2311'} + s_{31;32311'} + s_{23;232311'} +
  s_{32;1232311'} - s_{32;2311'} \right]
\nonumber
\\
& & - \ \left[ s_{12;1'} + s_{31;31'} + s_{23;231'} +
  s_{32;1231'} - s_{32;1'} \right]
\nonumber
\\
& & + \ \left[ s_{12;11'1'} +  s_{23;1'1'} +
  s_{31;311'1'}\right]
- 
\left[ s_{12;11'1'123} +  s_{23;1'1'123} +
  s_{31;311'1'123}\right]
\nonumber
\\
& & + \ {1 \over 2} \ [(\tau_{32}- \tau_{12})_{;2311'} + (\tau_{21} -
  \tau_{31})_{;32311'} + (\tau_{13} - \tau_{23})_{;232311'} 
- (\tau_{13} - \tau_{23})_{;1'1232311'} 
\nonumber
\\
& & + \ (\tau_{13} -  \tau_{23})_{;1'2311'} - (\tau_{32}-
  \tau_{12})_{;1'}
- (\tau_{21} -  \tau_{31})_{;31'} - (\tau_{13} -
  \tau_{23})_{;231'} 
\nonumber
\\
& & 
+ \ (\tau_{13} - \tau_{23})_{;1'1231'} + (\tau_{32}- \tau_{12})_{;11'1'} + (\tau_{21} -
  \tau_{31})_{;311'1'} - (\tau_{32}- \tau_{12})_{;11'1'123}
\nonumber
\\
& & - \ (\tau_{13} - \tau_{23})_{;1'1'123}
- (\tau_{21} -  \tau_{31})_{;311'1'123}]
\label{eq:37}
\end{eqnarray}
\noindent
with $U_2$, $U_3$ obtained by cycling the spacecraft indices. It can
readily be verified using equations (\ref{eq:26}, \ref{eq:27}) that
the laser noise remaining in this combination vanishes to first order
in the spacecraft relative velocities $\dot L_i$.

\subsection{The Beacon}

In order to derive the expression for $P_1$ let us consider the
following data combinations entering into the expression for $P$ given
in Section III
\begin{eqnarray}
\left[\eta_{23} + \eta_{32;1} + \eta_{13;1'1} - \eta_{13} \right]_{;3} & = & D_{3}
\left[D_{1} D_{1'} - I\right] D_{2'} \phi_1 \ ,
\label{eq:32a}
\\
\left[\eta_{32} + \eta_{23;1'} + \eta_{12,11'} - \eta_{12} \right]_{;2'} & = & D_{2'}
\left[D_{1'} D_{1} - I\right] D_{3} \phi_1 \ ,
\label{eq:32b}
\end{eqnarray}
where the expressions on the right-hand-sides follow from the chosen
order of the indices appearing on the left-hand-side of the above
equation. By applying our method we obtain
the final expression for $P_1$
\begin{eqnarray}
P_1 & = & D_{2'} \left[D_{1'} D_{1} - I \right] D_{3}
\left[\eta_{23;3} + \eta_{32;13} + \eta_{13;1'13} - \eta_{13;3}
\right]
\nonumber
\\
& & - D_{3} \left[D_{1} D_{1'} - I \right] D_{2'}
\left[\eta_{32;2'} + \eta_{23;1'2'} + \eta_{12,11'2'} - \eta_{12;2'}
\right]
\label{eq:33}
\\
& = & \left[\eta_{23;3311'2'} + \eta_{32;13311'2'} + \eta_{13;1'13311'2'} - \eta_{13;3311'2'}\right]
\nonumber
\\
& & - \left[\eta_{23;332'} + \eta_{32;1332'} + \eta_{13;1'1332'} - \eta_{13;332'}
\right]
\nonumber
\\
& & + \left[\eta_{32;2'2'3} + \eta_{23;1'2'2'3} + \eta_{12,11'2'2'3} - \eta_{12;2'2'3}
\right]
\nonumber
\\
& & - \left[\eta_{32;2'2'1'13} + \eta_{23;1'2'2'1'13} +
  \eta_{12,11'2'2'1'13} - \eta_{12;2'2'1'13} \right]
\label{eq:33a}
\end{eqnarray}
Equation (\ref{eq:33a}) can be rewritten in terms of the one-way
measurements $s_{ij}$, $\tau_{ij}$
\begin{eqnarray}
P_1 & = & \left[s_{23;3311'2'} + s_{32;13311'2'} + s_{13;1'13311'2'} - s_{13;3311'2'}\right]
\nonumber
\\
& & - \left[s_{23;332'} + s_{32;1332'} + s_{13;1'1332'} - s_{13;332'}
\right]
\nonumber
\\
& & + \left[s_{32;2'2'3} + s_{23;1'2'2'3} + s_{12,11'2'2'3} - s_{12;2'2'3}
\right]
\nonumber
\\
& & - \left[s_{32;2'2'1'13} + s_{23;1'2'2'1'13} +
  s_{12,11'2'2'1'13} - s_{12;2'2'1'13} \right]
\nonumber
\\
& & + \ {1 \over 2} \ [(\tau_{13} - \tau_{23})_{;3311'2'} 
- (\tau_{13} - \tau_{23})_{;1'13311'2'}
-  (\tau_{21} -  \tau_{31})_{;2'1'13311'2'} 
+ (\tau_{21} -  \tau_{31})_{;2'3311'2'}
\nonumber
\\
& & - \ (\tau_{13} - \tau_{23})_{;332'}
+ (\tau_{13} - \tau_{23})_{;1'1332'}
+ (\tau_{21} -  \tau_{31})_{;2'1'1332'}
- (\tau_{21} -  \tau_{31})_{;2'332'}
\nonumber
\\
& & + \ (\tau_{32}- \tau_{12})_{;11'2'2'3}
- (\tau_{32}- \tau_{12})_{;2'2'3}
- (\tau_{32}- \tau_{12})_{;11'2'2'1'13}
+ (\tau_{32}- \tau_{12})_{;2'2'1'13}]
\label{eq:34}
\end{eqnarray}
\noindent
with $P_2$, $P_3$ obtained by cycling the spacecraft indices in Eq.
(\ref{eq:34}). Substituting into equation (\ref{eq:34}) the laser
phase noise terms entering the $s_{ij}$ and $\tau_{ij}$, and applying
the expansion rules of equations (\ref{eq:25} - \ref{eq:27}), it can
again be shown that, to first order in the systematic relative
velocities of the spacecraft, laser phase noise is eliminated.

\subsection{The Monitor}

The derivation of the generalized ``Monitor'' combinations ($E_1, E_2,
E_3$) is more complicated, and rather different from the derivations
shown in the previous two subsections. One peculiarity of these
combinations is that they are not unique. It is indeed possible to
derive different expressions for each Monitor combination. These
combinations cancel the laser noises to the required order in the
velocities, and they differ only in the number of terms - delayed data
time series - they include.  We have derived expressions with $64$,
$32$, and $40$ $\eta$ terms (which we do not provide here). The
expression we present in this section shows the same number of
$\eta$-terms ($16$)  as $X_1$, $P_1$, and $U_1$.

\noindent
Let us consider the following terms entering into the expression for
$E$ derived in the previous section.
\begin{eqnarray}
\eta_{21;1'1} - \eta_{21} & = & \left[I - D_{1}D_{1'}\right] \phi_1 -
\left[I - D_{1}D_{1'} \right] D_{3'} \phi_2
\label{eq:38a}
\\
- \eta_{32,3'} - \eta_{23;1'3'} & = & D_{3'} \left[I - D_{1'}D_{1} \right] \phi_2 \ ,
\label{eq:38b}
\\
\eta_{31} - \eta_{31;11'} & = & - \left[I - D_{1'}D_{1} \right] \phi_1
+ \left[I - D_{1'}D_{1} \right] D_{2} \phi_3 \ ,
\label{eq:38c}
\\
\eta_{23;2} + \eta_{32;12} & = & - D_{2} \left[I - D_{1}D_{1'} \right]
\phi_3
\label{eq:38d}
\end{eqnarray}
If the delay operators were constant and commuted, adding these four
equations would cancel all laser phase noises and give $E$. Otherwise
the above expressions can be first combined in pairs
to remove the $\phi_2$, $\phi_3$ noises in two shear-free ways
\begin{eqnarray}
D_{3'} \left[I - D_{1'}D_{1} \right] \left[\eta_{21;1'1} -
  \eta_{21}\right]
& - & \left[I - D_{1}D_{1'} \right] D_{3'} \left[\eta_{32,3'} +
  \eta_{23;1'3'}\right]
\nonumber\\
& = & D_{3'} \left[I - D_{1'}D_{1} \right] \left[I -
  D_{1}D_{1'}\right] \phi_1
\label{eq:39a}
\end{eqnarray}
\begin{eqnarray}
D_{2} \left[I - D_{1}D_{1'} \right] \left[\eta_{31;11'}
- \eta_{31} \right]
& - & \left[I - D_{1'}D_{1} \right] D_{2} \left[\eta_{23;2} +
  \eta_{32;12}\right]
\nonumber\\
& = & D_{2} \left[I - D_{1}D_{1'} \right]\left[I - D_{1'}D_{1} \right]
\phi_1
\label{eq:39b}
\end{eqnarray}
Now we could of course repeat our iterative procedure by properly
using the delay operators shown on the right-hand-side of equations
(\ref{eq:39a}, \ref{eq:39b}), and derive the final expression for
$E_1$. However, this expression would include $64$ $\eta$-terms.
An alternative, and more elegant way to derive an expression for an
$E_1$ that has only $16$ $\eta$-terms is by noticing that if we first
apply inverse operators $D^{-1}_{3'}$ and $D^{-1}_{2}$ from equation
(\ref{eq:28}) to both sides of equation (\ref{eq:39a}) and
(\ref{eq:39b}) respectively, and then take the difference of the
resulting expressions, we get the following simpler expression for
$E_1$
\begin{eqnarray}
E_1 & \equiv &  
\left[I - D_{1'}D_{1} \right] 
\left[\eta_{21;1'1} - \eta_{21}\right]
- D^{-1}_{3'} \left[I - D_{1}D_{1'} \right] D_{3'} 
\left[\eta_{32,3'}  +  \eta_{23;1'3'}\right]
\nonumber
\\
& & 
+ \left[I - D_{1}D_{1'} \right] 
\left[\eta_{31} -  \eta_{31;11'}\right]
+ 
D^{-1}_{2} \left[I - D_{1'}D_{1} \right] D_{2} 
\left[\eta_{23;2} + \eta_{32;12}\right] \ ,
\nonumber
\\
& = & \left[\eta_{31} - \eta_{31;11'} - \eta_{31;1'1} + \eta_{31;11'1'1}\right]
- \left[\eta_{21} - \eta_{21;11'} - \eta_{21;1'1} + \eta_{21;1'111'}\right]
\nonumber
\\
& & + \left[\eta_{23;2} + \eta_{32;12} - \eta_{32;3'} - \eta_{23;1'3'}\right]
- \left[\eta_{23;2211' \bar{2}} + \eta_{32;12211'\bar{2}} - \eta_{32;3'3'1'1 \bar{3}'} \right.
\nonumber
\\
& & - \left. \eta_{23;1'3'3'1'1 \bar{3}'} \right] \ .
\label{eq:40}
\end{eqnarray}
In equation (\ref{eq:40}) we have introduced a bar over some
indices for representing the action of the corresponding inverse
operator.  It is easy to verify, to first order in the spacecraft
relative velocities, that the above expression is laser and optical
bench noise-free.

\noindent
Equation (\ref{eq:40}) can be recast in terms of the one-way
measurements $s_{ij}$ and $\tau_{ij}$
\begin{eqnarray}
E_1 & = & \left[s_{31} - s_{31;11'} - s_{31;1'1} + s_{31;11'1'1}\right]
- \left[s_{21} - s_{21;11'} - s_{21;1'1} + s_{21;1'111'}\right]
\nonumber
\\
& & + \left[s_{23;2} + s_{32;12} - s_{32;3'} - s_{23;1'3'}\right]
- \left[s_{23;2211' \bar{2}} + s_{32;12211'\bar{2}} - s_{32;3'3'1'1 \bar{3}'} \right.
\nonumber
\\
& & - \left. s_{23;1'3'3'1'1 \bar{3}'} \right] 
+ \ {1 \over 2} \ [(\tau_{32} - \tau_{12})_{;3'} 
- (\tau_{32} - \tau_{12})_{;3'1'1} 
- (\tau_{32} - \tau_{12})_{;3'11'} 
\nonumber
\\
& & + (\tau_{32} - \tau_{12})_{;3'1'111'} 
+ \ (\tau_{21} - \tau_{31})
- (\tau_{21} - \tau_{31})_{;11'}
- (\tau_{21} -  \tau_{31})_{;1'1}
\nonumber
\\
& & + (\tau_{21} -  \tau_{31})_{;11'1'1}
+ \ (\tau_{13}- \tau_{23})_{;2}
- (\tau_{13}- \tau_{23})_{;1'12}
- (\tau_{13}- \tau_{23})_{;2211' \bar{2}}
\nonumber
\\
& & + (\tau_{13}- \tau_{23})_{;1'12211' \bar{2}} ] \ ,
\label{eq:41}
\end{eqnarray}
\noindent
with $E_2$, $E_3$ obtained by cycling the indices.

\subsection{The $\zeta$ Combinations}

The expression for $\zeta_1$ derived in the previous section cancels
the laser noise exactly under the assumption of constant time delays.
Although perfect cancellation is no longer achieved when relative
motion between the spacecraft is included, the ordering of the delays
determined by our derivation of the expression for $\zeta_1$ given in
the previous section implies a minimization of the magnitude of the
remaining laser noises at least for the equilateral LISA case.

Consider the expression for $\zeta_1$ given in Eq. (\ref{eq:23}), now
however with semicolons rather then simple colons.  After some
algebra, it is possible to derive the leading order contribution due
to the residual laser noises remaining into $\zeta_1$:
\begin{equation}
\zeta_1 \simeq [\dot{\phi}_{2,4L} - \dot{\phi}_{2,3L}] \ 
({\dot L}_3 - {\dot L}_1) \ L 
+ [\dot{\phi}_{3,4L} - \dot{\phi}_{3,3L}] \ ({\dot L}_2 - {\dot L}_1)
\ L \ , 
\label{eq:44}
\end{equation}
where we have assumed the arm lengths to differ from a nominal LISA arm
length $L$ by only a few percents \cite{PPA98}.

This residual laser noise can be compared with the optical path and
proof mass noises in $\zeta_1$.  Using the derivative theorem for
Fourier transforms and taking the arm lengths to be the same, the
spectrum of the residual laser noise in $\zeta_1$ can be expressed in
terms of the spectrum of the raw laser phase noise, $S_{\phi}$, and
the velocities $\dot L_i$:

\begin{equation}
16 \ \pi^2 \ f^2 \ \sin^2(\pi f L) \ S_{\phi}(f) \
[(\dot L_2 - \dot L_1)^2 + (\dot L_3 - \dot L_1)^2 ] \ L^2
\end{equation}
From Section III and \cite{ETA00}, the spectrum of $\zeta_1$ due to
proof mass and optical path noises is equal to:
\begin{equation}
4 \ \sin^2(\pi f L) \ [24 \sin^2(\pi f L) \ S^{\rm proof \ mass}(f) 
+ 6 \ S^{\rm opt. \ path}(f)] \ .
\end{equation}
In Figure 2 we compare the spectrum of residual laser noise in
$\zeta_1$ and the optical path and proof mass noises in $\zeta_1$.
The parameters used were: $30$ Hz/$\sqrt{\rm Hz}$ for the raw laser
frequency fluctuations, $3 \times 10^{-15}$ m/sec$^2/\sqrt{\rm Hz}$
for the proof mass noise, and $20 \times 10^{-12}$ m/$\sqrt{\rm Hz}$
for aggregate optical path (shot noise, beam-pointing noise, etc.)
noise.  All the above spectra are one-sided. Figure 2 shows this
comparison using nominal $L = 16.67$ sec. arm lengths and
(pessimistically) taking the velocity differences to be $10$
m/sec.  From Figure 2, the residual laser noise in $\zeta_1$ for a
shearing array (but with the time delays applied as given in equation
(\ref{eq:23})) is $\simeq 7$ dB below the optical path and proof mass
noises.

\begin{figure}
\centering
\includegraphics[width=6.5in, angle=0.0]{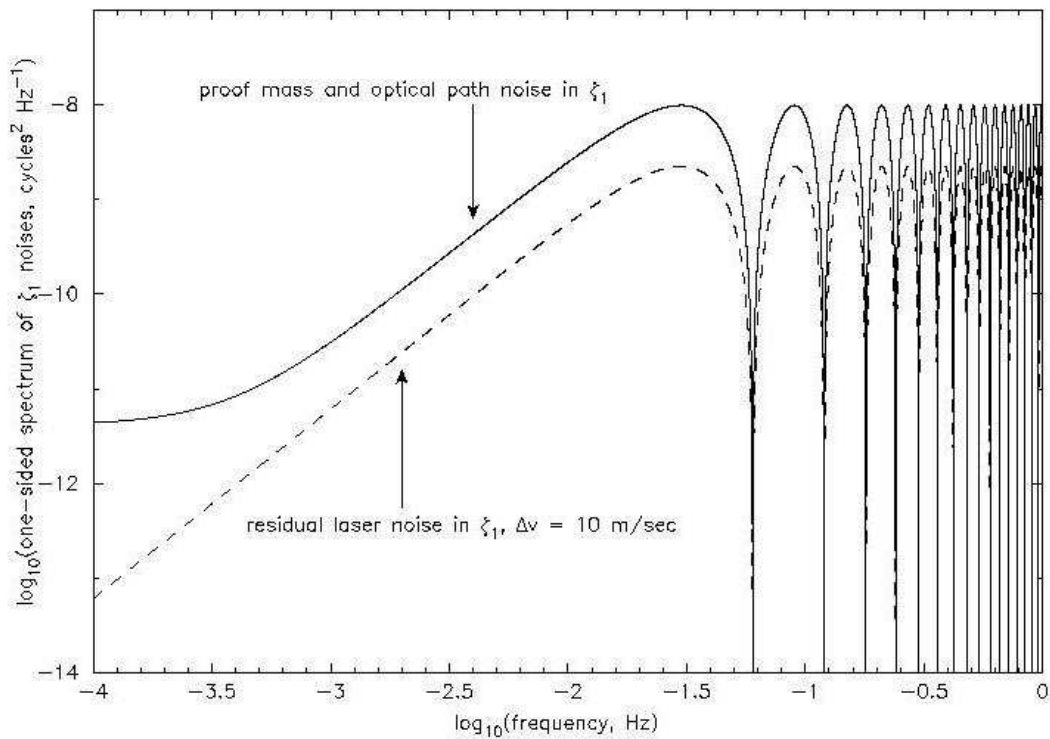}
\caption{
  Spectrum of proof mass and optical path noises in $\zeta_1$ compared
  with spectrum of residual laser noise, for a shearing array.
  Spectra are one-sided and expressed as (cycles)$^2$/Hz.  Parameters
  used: $30$ Hz/$\sqrt{{\rm Hz}}$ for the laser frequency
  fluctuations, $3 \times 10^{-15}$ m/sec$^2/\sqrt{{\rm Hz}}$ for the
  proof mass noise, $20 \times 10^{-12}$ m/$\sqrt{{\rm Hz}}$ for
  aggregate optical path (shot noise, beam-pointing noise, etc.)
  noise, $L = 16.67$ seconds, and the velocity differences have been
  taken to be equal to $10$ m/sec.  Laser noise does not cancel
  exactly in $\zeta_1$ for non-zero velocities, but is $\simeq$ 7 dB
  below optical path and proof mass noises.}
\label{fig:zeta_spectra}
\end{figure}

\section*{Acknowledgement}
\noindent
This research was performed at the Jet Propulsion Laboratory,
California Institute of Technology, under contract with the National
Aeronautics and Space Administration.

\end{document}